\documentstyle[aps,multicol,prb,epsf]{revtex}

\def\xiv{{\mbox{\boldmath{$\xi$}}}}

\newcommand{\nil}{\hspace*{0em}}
\begin{document}
\draft
\title{The spin-wave spectrum of the Jahn-Teller
system LaTiO$_{\mathbf{3}}$}
\author{Robert Schmitz,$^1$ Ora Entin-Wohlman,$^2$
Amnon Aharony,$^2$ A.\,Brooks Harris,$^3$ and Erwin
M\"uller-Hartmann$^1$}
\date{\today}
\address{$^1$Institut f\"ur Theoretische Physik,
Universit\"at zu K\"oln, Z\"ulpicher Stra{\ss}e 77, 50937 K\"oln,
Germany\\$^2$School of Physics and Astronomy, Raymond and Beverly
Sackler Faculty of Exact Sciences,
\\Tel Aviv University,
Tel Aviv 69978, Israel\\$^3$Department of Physics and Astronomy,
University of Pennsylvania, Philadelphia, Pennsylvania 19104, USA}
\maketitle

\begin{abstract}
We present an analytical calculation of the spin-wave spectrum of
the Jahn-Teller system LaTiO$_3$. The calculation includes all
superexchange couplings between nearest-neighbor Ti ions allowed
by the space-group symmetries: The isotropic  Heisenberg couplings
and the antisymmetric (Dzyaloshinskii-Moriya) and symmetric
anisotropies.  The calculated spin-wave dispersion has four
branches, two nearly degenerate  branches with small zone-center
gaps and two practically indistinguishable  high-energy branches
having large zone-center gaps. The two lower-energy modes are
found to be in satisfying agreement with neutron-scattering experiments.
In particular, the experimentally detected approximate isotropy in
the Brillouin zone and the small zone-center gap are well
reproduced by the calculations. The higher-energy branches have
not been detected yet by neutron scattering but their zone-center
gaps are in satisfying agreement with recent Raman data.\\

\end{abstract}

\begin{multicols}{2}

\section{Introduction}

The orthorhombic perovskite LaTiO$_3$ has long been considered as
a typical antiferromagnetic Mott insulator
($T_{\text{N}}=146\,\text{K}$). Albeit its rather small ordered
magnetic moment, 0.46$-0.57\mu_{B}$, \cite{keimer,cwik}
experimentally it seems not very different from a conventional
Heisenberg antiferromagnetic insulator. Indeed, the spin-wave
spectrum measured by Keimer {\it et al.} \cite{keimer} is well
described by a nearest-neighbor superexchange coupling having the
value 15.5 meV, accompanied by a weak ferromagnetic moment. The
latter has been attributed to  a small Dzyaloshinskii-Moriya
interaction, of about 1.1 meV. The experiment reported in Ref.~\onlinecite{cwik} has found that the antiferromagnetic order of
LaTiO$_3$ has a G-type structure along the crystallographic $a$
direction, while the ferromagnetic moment is along the $c$
direction.

Because of the unusually small ordered moment, it has been
proposed
 \cite{khaliullin} that perhaps the cubic Kugel-Khomskii Hamiltonian
\cite{KK} could be taken as a starting point for a successful
interpretation of LaTiO$_3$. However,  this cubic model has some
very unusual symmetries which inhibit the appearance of long-range
magnetic order at non-zero temperatures. \cite{harris03,harris04}
At strictly cubic symmetry, the fivefold degenerate $d$-levels on
the Ti ions are split by the crystal field of the oxygen octahedra
into the lower threefold degenerate $t_{2g}$ levels, (occupied in
Ti by a single electron) and the higher twofold degenerate $e_{g}$
levels. In real materials, those degeneracies are frequently
lifted by the Jahn-Teller distortion.

\begin{figure}[htb]
\leavevmode \epsfclipon \epsfxsize=7.truecm
\vbox{\epsfbox{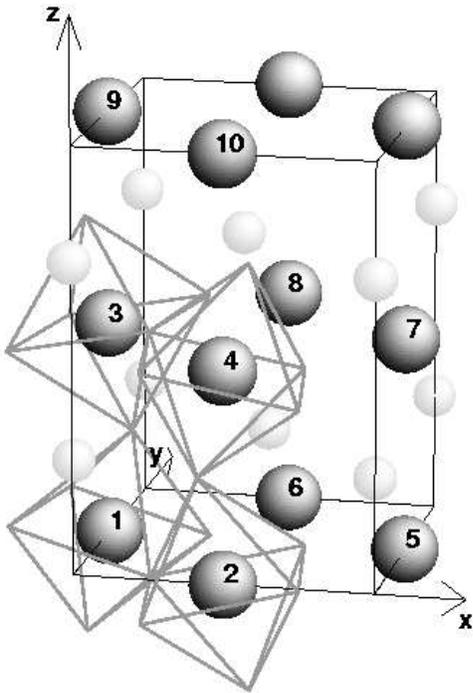}} \vspace{0.5cm} \caption{The
crystallographic structure of LaTiO$_3$. The ten Ti ions, which
constitute the twelve inequivalent nearest-neighbor Ti--Ti bonds
are enumerated. For simplicity, only the oxygen octahedra around
the four crystallographically inequivalent Ti sites are shown. La
ions from two layers are depicted as small spheres. We use
orthorhombic coordinates, in which the $x,y,z$ axes are oriented
along the crystallographic $a,b,c$ directions.}\label{bonds}
\end{figure}

Figure \ref{bonds} portrays the crystal structure of LaTiO$_3$
(the enumeration we use for the Ti sites is marked in the
figure). The unit cell contains four Ti ions, and the crystal has the symmetry
of the space group $Pbnm$. The Jahn-Teller effect in LaTiO$_3$ is
caused by the twisting of the Ti--O bonds with respect to each
other, i.\,e., by differences between the O--O bond lengths which
amounts to a deviation of certain  O--Ti--O bond angles away from
90$^{\circ}$. The distortion leads to a crystal field that splits
the levels,\cite{cwik} yielding a crystal-field gap of about
0.24\,eV between the orbitally non-degenerate ground state and the
first excited level, a value which has been confirmed by a study
of photo-electron spectroscopy. \cite{haverkort} A comparison of
the optical conductivity and of Raman data shows that the lowest
orbital excitation is centered at about 0.25\,eV.
\cite{grueninger} This value is in excellent agreement with the
estimate of the crystal-field splitting according to Ref.
\onlinecite{cwik}. Furthermore, the non-degenerate ground-state
orbital due to the crystal-field calculations given in Ref.
\onlinecite{cwik} is consistent with the orbital order found in
NMR measurements of the Ti--3$d$ quadrupole moment. \cite{kiyama}
The presence of orbital order at low temperatures has also been
inferred from measurements of the dielectric properties and the
dynamical conductivity. \cite{lunkenheimer}

\begin{figure}[htb]
\leavevmode \epsfclipon \epsfxsize=7.truecm
\vbox{\epsfbox{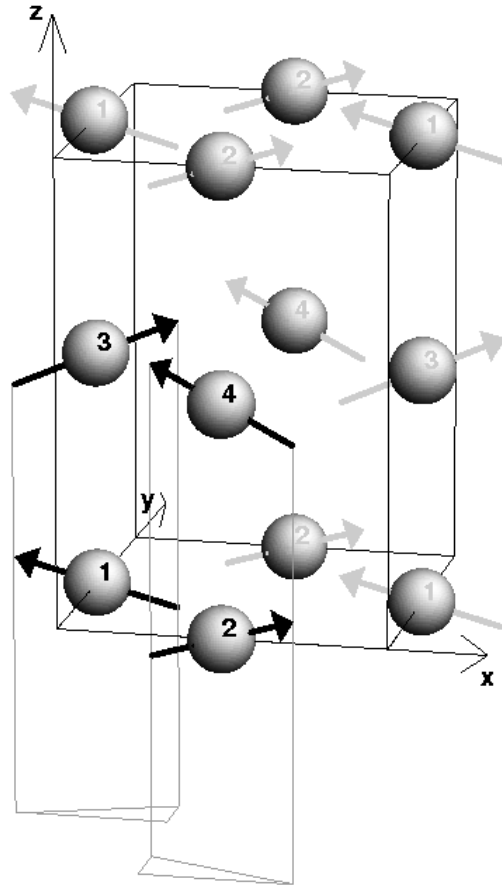}} \vspace{0.5cm}
 \caption{The magnetic order of the Ti ions in the classical
ground state of the effective spin Hamiltonian of the lattice. The
ions are enumerated according to the sublattice to which they
belong. }\label{cgsfig}
\end{figure}

An explanation of the magnetism of LaTiO$_{3}$, which is based on
the crystal-field calculation given in Ref.~\onlinecite{cwik}, is
presented in Ref.~\onlinecite{us}. The calculation included
spin-orbit interaction on the Ti ions as well, and found
accordingly that the superexchange coupling between neighboring Ti
ions consists of the isotropic Heisenberg exchange, and the
antisymmetric (Dzyaloshinskii-Moriya) and symmetric anisotropies,
which appear as a result of the spin-orbit interaction. These
anisotropies conspire together with the isotropic coupling to
determine the magnetic order at low temperatures, shown in Fig.
\ref{cgsfig}. By minimizing the magnetic energy of the classical
ground state it was found \cite{us} that the magnetic order of
LaTiO$_{3}$ is primarily that of a G-type antiferromagnet, with
the ordered moment along the crystallographic $a$ axis,
accompanied by a weak ferromagnetic moment along the $c$ axis, in
good agreement with experiment. In addition, it was found that
there is a small A-type moment of the spin components along the
$b$ axis, which (although not yet detected in experiment) is
allowed by the symmetry of the system.

In this paper we calculate  the spin-wave spectrum of LaTiO$_{3}$,
which evolves from the magnetic ground state found in Ref.
\onlinecite{us}. Since the magnetic unit cell contains four
sublattice magnetizations, the spin wave dispersion consists of
four branches. In the zero spin-orbit coupling limit, these four
branches collapse into two branches,  an acoustic mode and an
optical one, which both are two-fold degenerate. Accordingly, we term the 
two branches which evolve
from the (zero spin-orbit coupling) acoustic waves as `acoustic
modes', and those which evolve from the optical ones as `optical
modes'. At the Brillouin zone center, the energies of the two acoustic
branches do
not vanish but have gaps, of magnitudes 2.7\,meV and 3.0\,meV .
These values are quite close to the zone-center gap of about
3.3\,meV deduced from neutron scattering. \cite{keimer}
Furthermore, these two modes are approximately isotropic in the
Brillouin zone, again  in good agreement with the neutron
scattering experiment. \cite{keimer} We find that the two optical
modes are quasi-degenerate, having a zone-center gap of about
43.3\,meV. These modes have not been detected yet by neutron
scattering but are in good agreement with Raman data
\cite{grueninger} where at low temperatures an excitation peak is
seen, which is centered at about 40\,meV and which disappears at
$T_{\text{N}}$.

Our calculation employs linear spin-wave theory, which expresses
the deviations of the spins from their ground state configuration
in terms of  Holstein-Primakoff bosons. We therefore begin our
analysis by outlining in Sec.~II the determination of that ground
state configuration. We then continue to derive in Sec.~III the
spin-wave Hamiltonian, and to obtain the spin-wave dispersion.
Section IV contains a numerical study of the dispersion curves,
together with a detailed comparison with experiment. The summary
of our results is presented in Sec.~V.

\section{The magnetic ground state}

The analysis of the magnetic structure of LaTiO$_{3}$, carried out
in Ref.~\onlinecite{us}, involves several steps. First, a microscopic Hamiltonian containing the relevant
interactions on the Ti ions and  between nearest-neighbor Ti ions is derived.
Treating this Hamiltonian in perturbation theory, one then derives
the superexchange interactions between nearest-neighbor pairs of
spins of the electrons in the ground-state orbitals. This effective spin
Hamiltonian is summed over the entire Ti-lattice, to obtain the
magnetic Hamiltonian. Finally, one minimizes the resulting
magnetic energy of the system, to obtain the classical magnetic
ground state. In this section we briefly review these steps.

The derivation of the microscopic Hamiltonian starts from a
point-charge summation of the static crystal field for the Ti
ions, employing a full Madelung sum over the crystal. This
determines the eigenenergies and the eigenstates of the static
crystal field acting on each Ti ion, i.e., the crystal-field $d$
states. The effective hopping between the $d$ orbitals of
nearest-neighbor ions via the intervening oxygens is then written
in terms of a Slater-Koster parametrization of the Ti--O hopping.
The other interactions included in the microscopic Hamiltonian are
the on-site Coulomb interaction and the on-site
spin-orbit coupling on the Ti ions. In this way, the microscopic
Hamiltonian pertaining to a pair of nearest-neighbor Ti ions
(denoted $m$ and $n$) takes the form
\begin{eqnarray}
{\rm H}_{mn}={\rm H}^{0}_{mn}+{\rm V}_{mn}.\label{fullH}
\end{eqnarray}
Here
\begin{eqnarray}
{\rm H}_{mn}^{0}={\rm H}_{mn}^{\rm cf}+{\rm H}_{mn}^{\rm c},
\end{eqnarray}
where H$^{\rm cf}$ is the static crystal-field Hamiltonian, and
H$^{\rm c}$ describes the intra-ionic Coulomb correlations of a
doubly occupied $d$-shell. Because of the rather low symmetry of
the system, in treating the Ti$^{2+}$ ions which appear as
intermediate states of the exchange processes it is necessary to
take into account the full on-site Coulomb interaction matrix.
\cite{us} The other part of the Hamiltonian (\ref{fullH}) is
\begin{eqnarray}
{\rm V}_{mn}={\rm H}_{mn}^{\rm tun}+{\rm H}_{mn}^{\rm so},
\end{eqnarray}
in which H$^{\rm tun}$ is the kinetic energy, described in terms
of the effective hopping matrix, and H$^{\rm so}$ is the
spin-orbit interaction. This part is treated in perturbation
theory, in order to obtain from the  Hamiltonian (\ref{fullH}) an
effective spin Hamiltonian, pertaining to the spins of the two
Ti ions,  which acts within the Hilbert space of the fourfold
degenerate ground state of the unperturbed Hamiltonian ${\rm
H}^{0}$.

The detailed perturbation theory presented in Ref.
\onlinecite{us}, is carried out to second order in H$^{\rm tun}$
and up to  second order in the spin-orbit coupling (scaled by the
coupling strength $\lambda$). This procedure yields a rich
superexchange coupling between the spins of the non-degenerate
crystal-field ground states of the Ti$^{3+}$ ions. For a pair of
two nearest-neighbor Ti ions, the effective single-bond spin
Hamiltonian is found to be
\begin{eqnarray}
h_{mn}^{\nil}&=&J_{mn}^{\nil}{\mathbf{S}}_m^{\nil} \!\cdot\! 
{\mathbf{S}}_n^{\nil}+{\mathbf{D}}_{mn}^{\nil} \!\cdot\! 
\big({\mathbf{S}}_m^{\nil}\times{\mathbf{S}}_n^{\nil}\big)+
 {\mathbf{S}}_m^{\nil} \!\cdot\!  A_{mn}^{\text{s}} \!\cdot\! 
 {\mathbf{S}}_n^{\nil}, \nonumber \\ \label{hstart}
\end{eqnarray}
where $J_{mn}$ is the isotropic  Heisenberg coupling (second-order
in the tunnelling amplitudes, and independent of $\lambda$),
${\mathbf{D}}_{mn}$ is the Moriya vector (second-order in the
tunnelling amplitudes, and first order in $\lambda$), and
$A_{mn}^{\text{s}}$ is the symmetric anisotropy tensor
(second-order in the tunnelling amplitudes and in $\lambda$).  As
can be seen from Fig.~\ref{bonds}, there are 12 inequivalent
nearest-neighbor Ti--Ti bonds in the unit cell of LaTiO$_{3}$. By
the symmetry operations of the space-group $Pbnm$ the magnetic
couplings of all 8 intra-plane bonds can be expressed in terms of
those pertaining to the (12)-bond, and all 4 inter-plane ones in
terms of those of the bond (13). \cite{us} We
list the numerical values of the couplings in
Table \ref{microscres}. \cite{explanation} (The Tables are given on page
\pageref{microscres}.)

The magnetic Hamiltonian is found from the single-bond spin
Hamiltonian  (\ref{hstart}), by summing over the entire Ti
lattice. To this end, one decomposes the lattice into four
sublattices, corresponding to the four inequivalent Ti sites of
the unit cell (see Fig.~\ref{bonds}). Although all four sublattice
magnetizations are of equal magnitudes, their directions are all
different. Denoting the sublattice magnetization per site by
${\mathbf{M}}_i$, the macroscopic magnetic Hamiltonian is found to
be \cite{us}
\begin{equation}
H_{\text{M}}^{\nil}=\sum_{ij}\big[I_{ij}^{\hspace*{0em}}
{\mathbf{M}}_i^{\hspace*{0em}}\!\cdot \!{\mathbf{M}}_j^{\hspace*{0em}}
\!+\!{\mathbf{D}}_{ij}^{\text{D}}\!\cdot \!\big({\mathbf{M}}_i^{\hspace*{0em}}
\!\times\!{\mathbf{M}}_j^{\hspace*{0em}}\big)\!+\!
{\mathbf{M}}_i^{\hspace*{0em}}\!\cdot \!
 \Gamma_{ij}^{\nil}\! \cdot \!{\mathbf{M}}_j^{\hspace*{0em}}\big],\label{HM}
\end{equation}
where $ij$ runs over the sublattice pairs $(12),(13),(24),$ and
$(34)$ of Fig.~\ref{bonds}. Here, $I_{ij}$ are the macroscopic
isotropic couplings, ${\mathbf{D}}^{\text{D}}_{ij}$ are the
Dzyaloshinskii vectors (to leading order in the spin-orbit
coupling), which are the macroscopic antisymmetric anisotropies,
and $\Gamma_{ij}$ are the macroscopic symmetric anisotropy tensors
(of second order in the spin-orbit coupling). The relations
between those macroscopic couplings and the microscopic
single-bond couplings are listed in Table \ref{macrmicr}, and the
inter-relations between the macroscopic magnetic couplings of
different bonds, which are dictated by the space group symmetries,
are found in Table \ref{macsym}.

The minimization of the magnetic Hamiltonian (\ref{HM}) yields the
magnetic structure shown in Fig.~\ref{cgsfig}. Table \ref{cgs}
lists the details of this structure, in terms of the canting
angles $\varphi$ and $\vartheta$ according to Ref.
\onlinecite{us}. This structure  is going to be the basis for the spin-wave
expansion carried out in the next section.

\section{The spin-wave Hamiltonian}

The deviations of the spins away from their directions in the classical ground state may be  described in terms
of Holstein-Primakoff boson operators. In our case, the system
consists of four sublattices, which implies the introduction of
four different bosonic fields, and, in turn, four branches in the
spin-wave dispersion.

The first step in the standard calculation of spin-wave
dispersions is the rotation of the local coordinates at each
sublattice, $i$, such that the new $z$ axis will point in the
direction of the corresponding sublattice ground-state magnetization,
${\mathbf{M}}_i$. This rotation still leaves  the freedom to
choose the new local $x$ and $y$ axes, i.e., to rotate the new
coordinate system around its $z$ axis. Denoting the new local
coordinate system by $x'_{i}$, $y'_{i}$ and $z'_{i}$
($i=1,2,3,4$), we find that the convenient choice  for our
purposes (explained  in Appendix \ref{explcoordrot}) is
\begin{eqnarray}
\hat{z}'_{i}&=&\frac{{\mathbf{M}}_i}{M}, \ \
\hat{y}'_{i}=\frac{{\mathbf{M}}_{i}\times\hat{x}}{m_{i}}, \ \
\hat{x}'_{i}=\hat{y}'_{i}\times\hat{z}'_{i}, \nonumber\\
M&=&\big|{\mathbf{M}}_i\big|, \quad m_{i}=\sqrt{(M_{i}^{y})^{2}+(M_{i}^{z})^{2}}.\label{localcoor}
\end{eqnarray}
Consequently, there is a local rotation matrix $U_{i}$, pertaining
to each of the four sublattices,
\begin{eqnarray}
U_{i}=\left [\begin{array}{ccc}m_{i}&
-M_{i}^{y}M_{i}^{x}/m_{i}&-M_{i}^{z}M_{i}^{x}/m_{i}\\
0&M^{z}_{i}/m_{i}&-M_{i}^{y}/m_{i}\\
M^{x}_{i}&M^{y}_{i}&M^{z}_{i}\end{array}\right ],\label{Ui}
\end{eqnarray}
which rotates the orthorhombic into the local coordinate system.
We now apply this local rotation to the spin Hamiltonian
(\ref{hstart}), re-writing it for convenience  in  short-hand
notation
\begin{eqnarray}
h=\sum_{\langle mn\rangle}h_{mn}=\sum_{\langle
mn\rangle}{\mathbf{S}}_{m} \!\cdot\! A_{mn} \!\cdot\! {\mathbf{S}}_{n},\label{short}
\end{eqnarray}
where $A_{mn}$ is the 3$\times$3 superexchange matrix, comprising
all three types of magnetic couplings. In the rotated coordinate
system the spin Hamiltonian takes the form
\begin{eqnarray}
h=\sum_{\langle mn\rangle}{\mathbf{S}}_m' \!\cdot \! A_{mn}' \!\cdot \!{\mathbf{S}}_n',
\label{trafo}
\end{eqnarray}
where the primes denote  the rotated quantities,
\begin{eqnarray}
{\mathbf{S}}_m'=U_m^{\nil}\!\cdot \!{\mathbf{S}}_m^{\nil}, \;
A_{mn}'=U_m^{\nil}\!\cdot \!A_{mn}^{\nil}\!\cdot \! U_n^t.\label{trafo1}
\end{eqnarray}

We next introduce the Holstein-Primakoff boson fields \cite{hp} for each
of the four sublattices. Since we consider only the Ti ions, it is
convenient to use a coordinate system in which the Ti ions occupy
the sites of a simple cubic lattice, of unit lattice constant
(this picture is the appropriate one for comparing with the
experimental spin-wave data, \cite{keimer} as discussed in the
next section). It is also convenient to use a coordinate system
in which nearest-neighbor Ti ions are located along the axes
(namely, to rotate the orthorhombic coordinates by $-45^\circ$
around the $z$ axis, see Fig.~\ref{cgsfig}). Denoting the boson
fields of sublattice 1, 2, 3, and 4  by $a^{\nil}_{\bf R}$,
$b^{\nil}_{\bf R}$, $c^{\nil}_{\bf R}$, and $d^{\nil}_{\bf R}$,
respectively, where ${\bf R}$ is the radius vector to Ti No.~1 in
Fig.~ \ref{bonds}, the spin-wave Hamiltonian, in the harmonic
approximation takes the form
\begin{eqnarray}
h_{\rm SW} =h^{\text{sl}}_{12}+h^{\text{sl}}_{34}+
h^{\text{sl}}_{13}+h^{\text{sl}}_{24}. \label{BosLatHam}
\end{eqnarray}
The first term here refers to the lower plane of the unit cell,
the second to  the upper plane, and the last two terms refer to
the inter-plane couplings. Explicitly,
\end{multicols}
\begin{eqnarray}
h^{\text{sl}}_{12}= {\sum_{{\mathbf{R}}}}'\big\{&&
\big[C_{12}(1)+C_{16}(1)\big]\big(a_{{\mathbf{R}}}^{\dag}
a_{{\mathbf{R}}}^{\nil}
+b_{{\mathbf{R}}}^{\dag}b_{{\mathbf{R}}}^{\nil}
\big)\nonumber\\[-2ex]
&+&a_{{\mathbf{R}}}^{\dag}\big[C_{12}(2)
\big(b_{{\mathbf{R}}}^{\nil}+b_{{\mathbf{R}}-2{\mathbf{n}}_x}^{\nil}
\big)+C_{16}(2)\big(b_{{\mathbf{R}}-{\mathbf{n}}_x+{\mathbf{n}}_y}^{\nil}
+b_{{\mathbf{R}}-{\mathbf{n}}_x-
{\mathbf{n}}_y}^{\nil}\big)\big] \nonumber\\
&+&a_{{\mathbf{R}}}^{\dag}\big[C_{12}(3)\big(b_{{\mathbf{R}}}^{\dag}
+b_{{\mathbf{R}}-2{\mathbf{n}}_x}^{\dag}\big)+C_{16}(3)
\big(b_{{\mathbf{R}}-{\mathbf{n}}_x+{\mathbf{n}}_y}^{\dag}
+b_{{\mathbf{R}}-{\mathbf{n}}_x-
{\mathbf{n}}_y}^{\dag}\big)\big]\big\}+
\text{h.\,c.}, \nonumber\\
h^{\text{sl}}_{34}= {\sum_{{\mathbf{R}}}}'\big\{&&\big[C_{34}(1)
+C_{38}(1)\big]\big(c_{{\mathbf{R}}}^{\dag}
c_{{\mathbf{R}}}^{\nil} +
d_{{\mathbf{R}}}^{\dag}d_{{\mathbf{R}}}^{\nil}\big)\nonumber \\[-2ex]
&+&\;c_{{\mathbf{R}}}^{\dag}\big[C_{34}(2)\big(d_{{\mathbf{R}}}^{\nil}
+d_{{\mathbf{R}}-2{\mathbf{n}}_x}^{\nil}\big)
+C_{38}(2)\big(d_{{\mathbf{R}}
-{\mathbf{n}}_x+{\mathbf{n}}_y}^{\nil}+d_{{\mathbf{R}}-{\mathbf{n}}_x-
{\mathbf{n}}_y}^{\nil}\big)\big] \nonumber\\
&+&\;c_{{\mathbf{R}}}^{\dag}\big[C_{34}(3)\big(d_{{\mathbf{R}}}^{\dag}
+d_{{\mathbf{R}}-2{\mathbf{n}}_x}^{\dag}\big)+
C_{38}(3)\big(d_{{\mathbf{R}}-
{\mathbf{n}}_x+{\mathbf{n}}_y}^{\dag}+d_{{\mathbf{R}}-{\mathbf{n}}_x-
{\mathbf{n}}_y}^{\dag}\big)\big]\big\}+
\text{h.\,c.}, \nonumber\\
h^{\text{sl}}_{13}= {\sum_{{\mathbf{R}}}}'\big\{&&\;C_{13}(1)
\big(a_{{\mathbf{R}}}^{\dag}a_{{\mathbf{R}}}^{\nil}
+c_{{\mathbf{R}}}^{\dag}
c_{{\mathbf{R}}}^{\nil}\big)\nonumber\\[-2.5ex]
&+&\;
C_{13}(2)a_{{\mathbf{R}}}^{\dag}\big(c_{{\mathbf{R}}}^{\nil}+
c_{{\mathbf{R}}-2{\mathbf{n}}_z}^{\nil}\big)+
C_{13}(3)a_{{\mathbf{R}}}^{\dag}
\big(c_{{\mathbf{R}}}^{\dag}+c_{{\mathbf{R}}-
2{\mathbf{n}}_z}^{\dag}\big)\big\}+
\text{h.\,c.}, \nonumber\\
h^{\text{sl}}_{24}= {\sum_{{\mathbf{R}}}}'\big\{&&C_{24}(1)
\big(b_{{\mathbf{R}}}^{\dag}b_{{\mathbf{R}}}^{\nil}
+d_{{\mathbf{R}}}^{\dag}
d_{{\mathbf{R}}}^{\nil}\big)\nonumber\\[-2.5ex]
&+&\; C_{24}(2)b_{{\mathbf{R}}}^{\dag}
\big(d_{{\mathbf{R}}}^{\nil}
+d_{{\mathbf{R}}-2{\mathbf{n}}_z}^{\nil}\big)+C_{24}(3)
b_{{\mathbf{R}}}^{\dag}
\big(d_{{\mathbf{R}}}^{\dag}+d_{{\mathbf{R}}-
2{\mathbf{n}}_z}^{\dag}\big)\big\}+
\text{h.\,c.}.\label{BosLatHam1}
\end{eqnarray}
Here ${\mathbf{n}}_{\alpha}$ is a unit vector along the
$\alpha$-direction, where $\alpha =x, y,z$, and the Ti ion marked
by 1 in Fig.~\ref{bonds} is at the origin. The summations then
extend only over the Ti ions No.~1 in each unit cell (this is indicated
by the prime on the summation symbols). The coupling coefficients
$C_{mn}(\ell )$ in Eq.~(\ref{BosLatHam1}) are given by
combinations of the superexchange matrix elements
$(A'_{mn})^{\alpha\beta}$,
\begin{eqnarray}
C_{mn}^{\nil}(1)&=&-\,\mbox{$\frac{1}{2}$}(A_{mn}')^{zz}, \nonumber  \\
C_{mn}^{\nil}(2)&=&\quad \mbox{$\frac{1}{4}$}\big[(A_{mn}')^{xx}+
(A_{mn}')^{yy}+i\big((A_{mn}')^{yx}-(A_{mn}')^{xy}\big)\big], \nonumber  \\
C_{mn}^{\nil}(3)&=&\quad \mbox{$\frac{1}{4}$}\big[(A_{mn}')^{xx}-
(A_{mn}')^{yy}+i\big((A_{mn}')^{yx}+(A_{mn}')^{xy}\big)\big].
\label{BosSBHam1}
\end{eqnarray}
In writing down Eq.~(\ref{BosLatHam1}), we have omitted constant
terms. The transformation to the Holstein-Primakoff operators
yields also terms which are linear in the boson fields; these
vanish upon summing over all bonds (see Appendix
\ref{explcoordrot}) due to the proper choice of the local coordinate system.

Our magnetic unit cell is spanned by the vectors $(1,1,0)$,
$(1,-1,0)$, and $(0,0,2)$,  and the corresponding magnetic Brillouin zone
(MBZ) is defined by
\begin{eqnarray}
|q_{x}+q_{y}|\leq \pi, \ \ \ |q_{z}|\leq \frac{\pi}{2}.\label{BZ}
\end{eqnarray}
By introducing the Fourier transforms of the operators,
\begin{eqnarray}
a_{{\mathbf{R}}}^{\dag}&=&\mbox{$\sqrt{\frac{1}{N}}$}\sum_{{\mathbf{q}}\in
\text{MBZ}}e^{i{\mathbf{q}}{\mathbf{R}}}a_{{\mathbf{q}}}^{\dag},
\quad
b_{{\mathbf{R}}}^{\dag}=\mbox{$\sqrt{\frac{1}{N}}$}\sum_{{\mathbf{q}}\in
\text{MBZ}}e^{i{\mathbf{q}}({\mathbf{R}}+{\mathbf{n}}_{x})}
b_{{\mathbf{q}}}^{\dag},\nonumber\\
c_{{\mathbf{R}}}^{\dag}&=&\mbox{$\sqrt{\frac{1}{N}}$}\sum_{{\mathbf{q}}\in
\text{MBZ}}e^{i{\mathbf{q}}({\mathbf{R}}+{\mathbf{n}}_{z})}
c_{{\mathbf{q}}}^{\dag},\quad
d_{{\mathbf{R}}}^{\dag}=\mbox{$\sqrt{\frac{1}{N}}$}\sum_{{\mathbf{q}}\in
\text{MBZ}}e^{i{\mathbf{q}}({\mathbf{R}}+{\mathbf{n}}_{x}+{\mathbf{n}}_{z})}
d_{{\mathbf{q}}}^{\dag},
\end{eqnarray}
where $N$ is the total number of magnetic unit cells, the
spin-wave Hamiltonian (\ref{BosLatHam}) becomes
\begin{eqnarray}
h_{\rm SW}=\sum_{\bf q}h_{\rm SW}({\bf q}),\label{HSW}
\end{eqnarray}
where
\begin{eqnarray}
h_{\rm SW}({\bf q})&=&C_{1}\Bigl (a^{\dagger}_{\bf q}a^{\nil}_{\bf
q}+b^{\dagger}_{\bf q}b^{\nil}_{\bf q}+c^{\dagger}_{\bf
q}c^{\nil}_{\bf
q}+d^{\dagger}_{\bf q}d^{\nil}_{\bf q}\Bigr )\nonumber\\
&+&\Bigl [C^{\parallel}_{2}(\cos q_{x}+\cos q_{y})\Bigl
(a^{\dagger}_{\bf q}b^{\nil}_{\bf q}+c^{\dagger}_{\bf
q}d^{\nil}_{\bf q}\Bigr )+\text{h.\,c.}\Bigr ] +\Bigl [C^{\perp}_{2}\cos
q_{z}\Bigl (a^{\dagger}_{\bf q}c_{\bf
q}+b^{\dagger}_{\bf q}d_{\bf q}\Bigr )+\text{h.\,c.}\Bigr ]\nonumber\\
&+&\Bigl [(C^{\parallel}_{3}\cos q_{x}+C^{\parallel *}_{3}\cos
q_{y})\Bigl (a^{\dagger}_{\bf q}b^{\dagger}_{-{\bf
q}}+c^{\dagger}_{\bf q}d^{\dagger}_{-{\bf q}}\Bigr )+\text{h.\,c.}\Bigr ]
+\Bigl [C^{\perp}_{3}\cos q_{z}a^{\dagger}_{\bf
q}c^{\dagger}_{-{\bf q}}+C_{3}^{\perp *}\cos q_{z}b^{\dagger}_{\bf
q}d^{\dagger}_{-{\bf q}}+\text{h.\,c.}\Bigr ].\label{start}
\end{eqnarray}
The coefficients appearing in this equation are linear
combinations of the previous coefficients $C_{mn}(\ell )$ (see
Appendix \ref{explcoordrot}),
\begin{eqnarray}
C_1^{\nil}&=&2C_{13}^{\nil}(1)+4C_{12}^{\nil}(1)=C_1^{*}, \quad
C_2^{\perp }=2C_{13}^{\nil}(2)=C_2^{\perp *}, \quad
C_2^{\|}=2C_{12}^{\nil}(2), \quad C_3^{\perp }=2C_{13}^{\nil}(3),
\quad C_3^{\| }=2C_{12}^{\nil}(3). \label{swcoeff}
\end{eqnarray}
These are related to the original spin-coupling coefficients of
Eq.~(\ref{hstart}), but  are  not reproduced here explicitly,
since their expressions are very long.

The spin-wave dispersion pertaining to the Hamiltonian (\ref{HSW})
is calculated in Appendix \ref{diag}, leading to the result
\begin{eqnarray}
\Omega _{1}^{2}({\bf q})&=&(C_{1}-C_{2}^{\perp}\cos q_{z})^{2}
-|C_{3}^{\perp}|^{2}\cos^{2}q_{z}+|C_{2}^{\parallel}|^{2}(\cos
q_{x}+\cos q_{y})^{2}\nonumber\\
&-&|C^{\parallel}_{3}\cos q_{x}+C^{\parallel *}_{3}\cos
q_{y}|^{2}-(\cos q_{x}+\cos q_{y})W(\cos q_{z}),\nonumber\\
\Omega_{2}^{2}({\bf q})&=&\Omega_{1}^{2}({\bf q}+{\bf Q}),\ \ \
{\rm with}\ \ {\bf Q}=(0,0,\pi ),\nonumber\\
\Omega_{3}^{2}({\bf q})&=&\Omega_{1}^{2}({\bf q}+{\bf Q}'),\ \ \
{\rm with}\ \ {\bf Q}'=(\pi,\pi,0 ),\nonumber\\
\Omega_{4}^{2}({\bf q})&=&\Omega_{1}^{2}({\bf q}+{\bf Q}''),\ \ \
{\rm with}\ \ {\bf Q}''={\bf Q}+{\bf Q}'=(\pi,\pi,\pi
),\label{omegas}
\end{eqnarray}
where
\begin{eqnarray}
W^{2}(\cos q_{z})&=&4\Bigl [(C_{1}-C_{2}^{\perp}\cos
q_{z})^{2}-|C^{\perp}_{3}|^{2}\cos^{2}q_{z}\Bigr ]\Bigl
[|C_{2}^{\parallel}|^{2}-\Bigl
(\frac{C_{3}^{\parallel}+C_{3}^{\parallel *}}{2}\Bigr )^{2}\Bigr
]\nonumber\\
&+&\Bigl [(C^{\perp
*}_{3}C^{\parallel}_{2}+C^{\perp}_{3}C^{\parallel *}_{2})\cos
q_{z}+(C_{1}-C_{2}^{\perp}\cos
q_{z})(C^{\parallel}_{3}+C^{\parallel *}_{3})\Bigr ]^{2}.
\end{eqnarray}
\begin{multicols}{2}
\noindent Each of the branches has tetragonal symmetry, i.e.,
$\Omega_i(q_x,q_y,q_z)=\Omega_i(q_y,q_x,q_z)=
\Omega_i(-q_x,q_y,q_z)=\Omega_i(q_x,-q_y,q_z)=\Omega_i(q_x,q_y,-q_z)
$.

Equations (\ref{omegas}) contain our final result for the
spin-wave spectrum of LaTiO$_{3}$. Evidently, the details of the
spectrum can be obtained only numerically: One has to write the
spin-wave coefficients, Eqs.~(\ref{swcoeff}), in terms of those
appearing in Eqs.~(\ref{BosSBHam1}),  and express the latter via
Eqs.~(\ref{short}) and (\ref{trafo1}) in terms of the original
coefficients of the spin Hamiltonian (\ref{hstart}) using the
values listed in Table \ref{microscres}. These results are then
used in constructing the dispersion. We carry out this procedure
in the next section, confining ourselves to the wave vectors
explored in the  neutron scattering and Raman experiments,
respectively.

When the spin-orbit coupling $\lambda$ is set to zero, the
coefficients appearing in Eqs.~(\ref{omegas}) simplify to
\begin{eqnarray}
C_{1}&=&2J_{12}+J_{13}, \ \ \
C^{\perp}_{2}=C^{\parallel}_{2}=0,\nonumber\\
C^{\perp}_{3}&=&-J_{13},\ \ \ C^{\parallel}_{3}=-J_{12},
\end{eqnarray}
where $J_{12}$ is the isotropic in-plane Heisenberg coupling, and
$J_{13}$ is the Heisenberg coupling between planes. In that case
[note that $\cos q_{z}=|\cos q_{z}|$ in the Brillouin zone, see
Eq.~(\ref{BZ})]
\begin{eqnarray}
\Omega_{1}^{2}({\bf q})&=&\Omega_{2}^{2}({\bf q}) =
(2J_{12}+J_{13})^{2}\nonumber\\
&-&\bigl (J_{12}(\cos q_{x}+\cos q_{y})+J_{13}|\cos q_{z}|\bigr
)^{2},
\end{eqnarray}
while the expression for $\Omega_{3}^{2}({\bf q})=\Omega_{4}^{2}({\bf q})$ is
obtained upon changing $\cos q_{x}+\cos q_{y}$ to $-\cos
q_{x}-\cos q_{y}$. At the zone center $\Omega_{1}$ and
$\Omega_{2}$ vanish, while $\Omega_{3}$ and $\Omega_{4}$ have a
gap equal to $\sqrt{8J_{12}J_{13}}$. Obviously, in the absence of
spin-orbit coupling the magnetic unit cell includes only two
sublattices (in that case, sublattice 1 and sublattice 4 can be combined into one sublattice, and so can sublattice 2 and sublattice 3 in Fig.
\ref{cgsfig}). The Brillouin zone corresponding to this smaller
magnetic cell is twice as large as the one of Eq.~(\ref{BZ}). By
``folding out" the optical mode into this larger Brillouin zone,
one reproduces the usual gap-less dispersion of the pure
Heisenberg model.  At finite values of the spin-orbit coupling all
modes have gaps at the zone center, but those of $\Omega_{1}$ and
$\Omega_{2}$ are much smaller than the ones of the other two
modes. For this reason, we term the $\Omega_{1}({\bf q})$ and the
$\Omega_{2}({\bf q})$ branches `acoustic modes' and
$\Omega_{3}({\bf q})$ and $\Omega_{4}({\bf q})$ are referred to as
optical modes. Optical spin-wave modes have been detected, for
instance, in  bilayer cuprates. \cite{reznik,hayden,pailhes}

\section{Numerical results for the spin-wave dispersion}

For the model parameters we use, it turns out that the two
acoustic branches as well as the two optical branches are nearly degenerate. The reason is the smallness of the
angle $\varphi$, which leads to an additional translational symmetry which is
nearly fulfilled by  the classical ground state. This
``quasi"-symmetry corresponds to the translation by the vector
${\mathbf{R}}_{14}$ which connects the Ti ions No.~1 and 4 (see
Figs.~\ref{bonds} and \ref{cgsfig}). For $\varphi$=0 this
symmetry is exact, and the magnetic unit cell contains only two
ions. In that case the spin-wave dispersion consists of two
branches. As we have a small deviation from this ideal case, we
obtain two pairs of quasi-degenerate  branches.

\subsection{Comparison of the acoustic branches
with neutron scattering data}

We begin our discussion here by recalling the experimental
results of Ref.~\onlinecite{keimer}. The authors of Ref.~\onlinecite{keimer} have fitted their neutron scattering data
with an isotropic single-branch spectrum parametrized as
\begin{eqnarray}
\Omega({\mathbf{q}}) \simeq
J\sqrt{\left(3+\frac{\Delta^2}{6J^2}\right)^2\!\!-\!(\cos q_x+\cos
q_y+\cos q_z)^2}. \!\!\label{expfit}
\end{eqnarray}
This assumes an isotropic
Heisenberg coupling, $J$, for the entire Ti lattice, namely, the
same coupling for the bond (12) and the bond (13) of Fig.
\ref{bonds}, and introduces  a zone-center spin-wave
gap, $\Delta$. The experimentally determined values of these
parameters are
\begin{equation}
J=15.5 \pm 1.0 \,\text{meV}, \quad \Delta=3.3 \pm 0.3\,\text{meV}.
\label{expval}
\end{equation}
In the following, we compare the fitted function,  Eq.
(\ref{expfit}), with the acoustic branches
$\Omega_1({\mathbf{q}})$ and  $\Omega_2({\mathbf{q}})$.

Although the symmetry of our spin-wave Hamiltonian allows for two
acoustic modes,  the resolution of the dispersion measurements,
which amounts to about 10\,\% at any given point ${\mathbf{q}}$ in
the Brillouin zone, \cite{keimerprivat} is insufficient to
resolve the two  branches. To demonstrate this point,
and to compare in detail the experimental findings with our
expressions, we proceed as follows. Firstly, we average the
Heisenberg couplings pertaining to the different bonds (calculated
in Ref.~\onlinecite{us}) over the six Ti--Ti bonds in which each
Ti ion is participating,
\begin{equation}
\frac{4J_{12}+2J_{13}}{6}=15.89 \,\text{meV}.
\end{equation}
Clearly this value agrees with the experimental one given in Eq.
(\ref{expval}), within the  accuracy of the experiment.  Secondly,
we calculate the zone-center gaps as found from our calculation.
Following the numerical procedure outlined at the end of the
previous section, we find
\begin{eqnarray}
\Delta_1&=&\Omega_1({\mathbf{0}})=2.71 \,\text{meV},\nonumber\\
\Delta_2&=&\Omega_2({\mathbf{0}})=2.98 \,\text{meV}.\label{gap}
\end{eqnarray}
We have found that the splitting between the two calculated
acoustic branches reaches its maximum at the zone center, where
\begin{equation}
\frac{\Delta_1}{\Delta_2}=91.14\,\%.
\end{equation}
This discrepancy is within the uncertainty of about 10\,\% of the
measured spin-wave energies of Ref.~\onlinecite{keimer}.

Away from the zone center the two acoustic branches are
quasi-degenerate. We estimate the tetragonal anisotropy of the
dispersion by comparing the dispersions at wave vectors
${\mathbf{q}}=(\pi/2,0,0)$ and ${\mathbf{q}}=(0,0,\pi/2)$,
\begin{equation}
\frac{\Omega_1(0,0,\mbox{$\frac{\pi}{2}$})}
{\Omega_1(\mbox{$\frac{\pi}{2}$},0,0)}=91.34\,\%,
\quad
\frac{\Omega_2(0,0,\mbox{$\frac{\pi}{2}$})}
{\Omega_2(\mbox{$\frac{\pi}{2}$},0,0)}=91.29\,\%. \label{tetraniso}
\end{equation}
This implies that the tetragonal anisotropy is also less than the
uncertainty of the measured spin-wave energies. The calculated
dispersions along selected directions in the Brillouin zone are
depicted in Figs.~\ref{disp}, together with the optical branches
which we will discuss in Sec.~\ref{Raman} and the experimental
dispersion computed from Eq.~(\ref{expfit}). The agreement between
the acoustic branches and the experimental dispersion is satsifying.

It is harder to infer  the experimentally quoted value
\cite{keimer} of the Dzyaloshinskii-Moriya interaction, 1.1\,meV,
(which does not agree well with our values for the Dzyaloshinskii
vectors, see Table \ref{microscres}), from the calculated
dispersion. We therefore attempt to estimate the effects of the
two types of anisotropies, antisymmetric and symmetric, on the
spin-wave dispersion by analyzing two cases: (i)  Switching off
all antisymmetric anisotropies,
${\mathbf{D}}^{\nil}_{mn}={\mathbf{0}}$, (all other terms are
accounted for according to their calculated values, see Table
\ref{microscres}) and (ii) switching off all symmetric
anisotropies, $A^{\text{s}}_{mn}=0$, while keeping the
contributions of the antisymmetric ones. In both cases we examine
the spin canting, i.e., the ground-state configuration of the
magnetization, and the zone-center gap of the dispersion. (The
dispersion away from the zone center is dominated by the
Heisenberg couplings.)

(i) In the absence of   the Dzyaloshinskii-Moriya interaction, the
canting practically disappears. We find that the canting angles
almost vanish,
\begin{equation}
\varphi=-0.04^\circ, \; \vartheta=0.00^\circ ,\; \text{if }
{\mathbf{D}}^{\nil}_{mn}={\mathbf{0}}.
\end{equation}
However, the zone-center gap   is  enhanced compared to its actual
values,  Eq.~(\ref{gap}),
\begin{equation}
\Delta_1=\Delta_2=4.73\,\text{meV}, \; \text{if }
{\mathbf{D}}^{\nil}_{mn}={\mathbf{0}}.
\end{equation}

(ii) In the absence of the symmetric anisotropies the spin canting
is almost the same as given in Table \ref{cgs},
\begin{equation}
\varphi=1.47^\circ, \; \vartheta=0.80^\circ ,\;  \text{if
}A^{\text{s}}_{mn}=0.
\end{equation}
Switching off continuously  the symmetric anisotropies, we find
that the zone-center gap first closes and then even becomes
imaginary as the symmetric anisotropies approach zero. This
unphysical result shows that one is not allowed to consider only
the antisymmetric anisotropies resulting from the spin-orbit
interaction, without including the symmetric ones as well. Indeed,
as has been already pointed out in Refs. \onlinecite{kaplan} and
\onlinecite{shekht}, a systematic treatment of the effect of the
spin-orbit interaction on the spin couplings must include both
anisotropies. They both contribute to the magnetic energy terms of
the same order in the spin-orbit coupling parameter.

Comparing these two fictitious cases, we conclude that the
spin-canting is dominated by the antisymmetric anisotropies, while
the zone-center gap of the dispersion is governed by the symmetric
anisotropies. It is therefore a somewhat  questionable procedure
to deduce the antisymmetric anisotropy of the spin coupling from
the spin-wave dispersion, taking into account  only the Moriya
vectors, as has been done in Ref.~\onlinecite{keimer}. This is
again related to the fact that both the Dzyaloshinskii-Moriya
interactions and the symmetric anisotropies induced by the
spin-orbit coupling appear in the same order in the magnetic
energy and in the spin-wave dispersion. \cite{shekht1}

The manner by which the various anisotropic spin couplings, in a
low-symmetry system like LaTiO$_3$, can be deduced from an
experimentally obtained spin-wave spectrum therefore remains
unsettled. In our case, the spin-wave Hamiltonian, Eq.
(\ref{start}), depends on 8 parameters [note that some of the
coefficients, Eqs.~(\ref{swcoeff}), are complex]. Furthermore,
even the knowledge of these 8 parameters does not suffice in our
case to trace backwards the parameters of the spin Hamiltonian,
Eq.~(\ref{hstart}). The reason being that the coefficients
involving the matrix elements ${(A'_{mn})}^{xz}$ and
${(A'_{mn})}^{yz}$ [see Eq.~(\ref{BosSBHam1})] disappear
altogether from the spin-wave Hamiltonian (see Appendix
\ref{explcoordrot}). The conclusion is that it is possible to use
certain numerical values for the various types of spin couplings
and to investigate their consistency with the experimentally
detected spin-wave dispersion (as done above). However, an unequivocal
deduction of spin-coupling parameters from spin-wave spectra is not possible
due to the low symmetry of this system.

\subsection{The optical branches} \label{Raman}

\begin{figure}
\begin{center}
\hspace*{1cm}(a) ${\mathbf{q}}=\frac{\pi}{2}(1,1,1)\xi$\\[2ex]
\end{center}
\leavevmode \epsfclipon \epsfxsize=7.9truecm
\vbox{\epsfbox{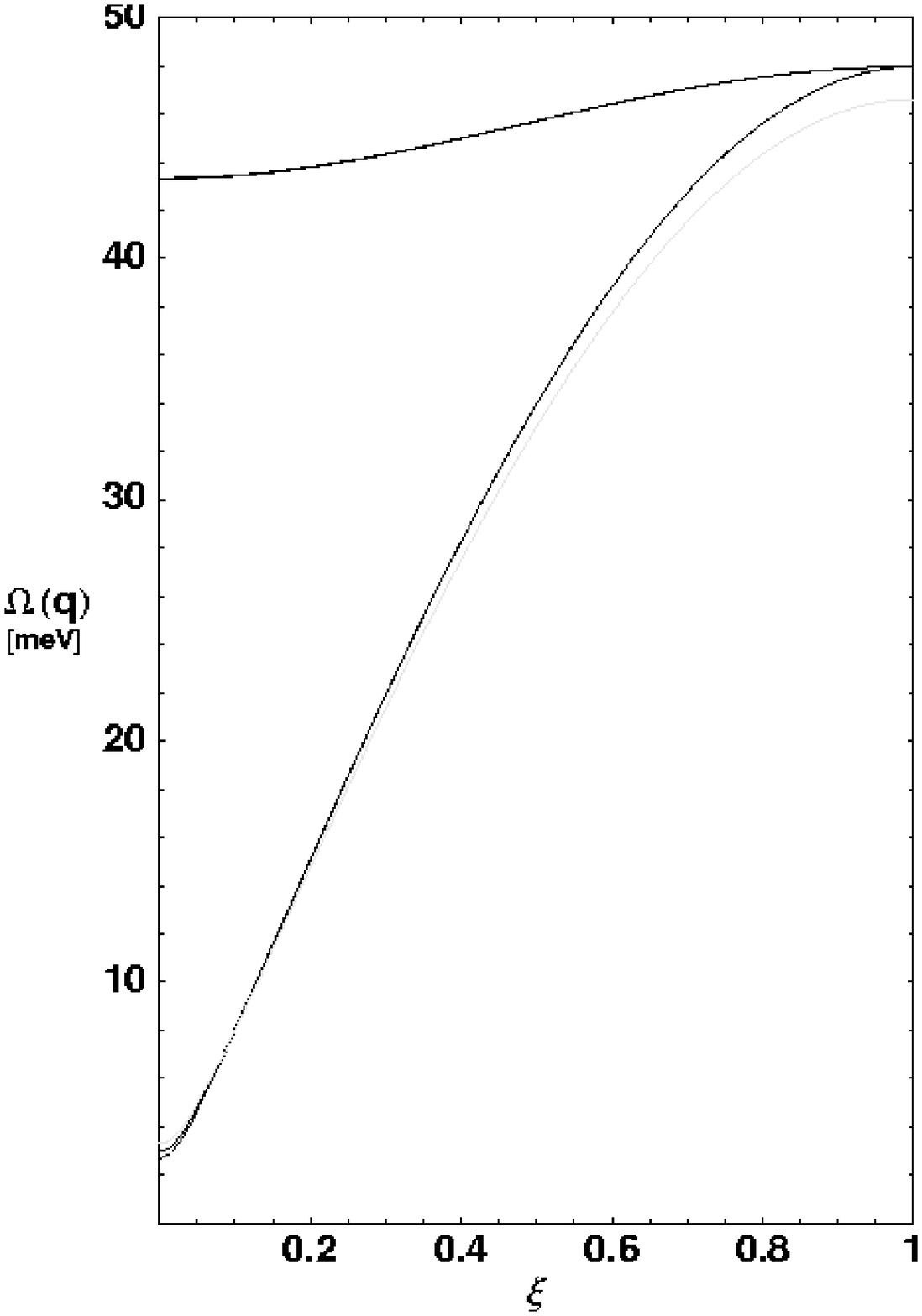}} \vspace{0.5cm} %\vspace*{10cm}
\caption{The spin-wave
dispersion along selected directions in the magnetic Brillouin zone. We use
pseudo-cubic coordinates, in which the Ti ions No.~1 and 2 are
located along the $x$ axis. Panels (a)--(c) show (in black) the
four branches $\Omega_i({\mathbf{q}})$ of the calculated
dispersion and (in grey) the single branch $\Omega({\mathbf{q}})$
which has been fitted onto neutron scattering experiments,
Eq.~(\ref{expfit}). The acoustic branches $\Omega_1({\mathbf{q}})$ and
$\Omega_2({\mathbf{q}})$ are quasi-degenerate, such that away from
the zone center no splitting between them can be seen.  The
optical branches $\Omega_3({\mathbf{q}})$ and
$\Omega_4({\mathbf{q}})$ are practically indistinguishable over
the entire Brillouin zone. (a) The dispersion along (1,1,1). This
direction is chosen because the experimental paper on the neutron
scattering contains a plot along this direction where the measured
points of the dispersion are shown. \cite{keimer} Though the
calculated acoustic branches of the dispersion give slightly lower energies at the
zone-center than the fitted function and
slightly higher energies at the zone edge, these deviations are
within the uncertainty of the measurement and hence, the agreement
of our calculated dispersion with the the measured points and with
the fitted function is satisfying. The splitting of the calculated
acoustic branches at the zone center is too small to be resolved
in the experiment. From panels (b) and (c) one can see that the
tetragonal anisotropy of the calculated acoustic branches is rather small.
The agreement between the acoustic branches and the neutron
scattering data is satisfying also along the  (1,0,0) [panel (b)] and
(0,0,1) [panel(c)] directions.} \label{disp}
\end{figure}

The two calculated optical branches, depicted in Figs.~\ref{disp}, are
practically indistinguishable in the entire Brillouin zone. Their
zone-center gaps are
\begin{eqnarray}
\Delta_3&=&\Omega_3({\mathbf{0}})=43.32 \,\text{meV},\nonumber\\
\Delta_4&=&\Omega_4({\mathbf{0}})=43.34
\,\text{meV}.\label{optgap}
\end{eqnarray}

So far, branches with such a large zone-center gap have not been
detected by neutron scattering. \cite{keimer} Possible reasons
are: (i) The signal in the energy range of the optical branches
has a rather low intensity (as compared to the lower energy
regions); (ii) The spin-wave signal in this energy range is
accompanied, and possibly is hidden, by phonon excitations.
\cite{keimerprivat} However, despite of these two problems, in
principle it might be possible to detect the dispersion of the
optical branches by neutron scattering. \cite{keimerprivat} Our
prediction is that the dispersion of the optical modes will be
qualitatively different from that of the acoustic ones. These
modes will not have the approximate isotropy of the acoustic
modes, but will show a larger tetragonal anisotropy. We find
\begin{equation}
\frac{\Omega_3(0,0,\mbox{$\frac{\pi}{2}$})}
{\Omega_3(\mbox{$\frac{\pi}{2}$},0,0)}=70.47\,\%, \quad
\frac{\Omega_4(0,0,\mbox{$\frac{\pi}{2}$})}
{\Omega_4(\mbox{$\frac{\pi}{2}$},0,0)}=70.44\,\%.
\end{equation}
These relations can serve as a further check of our model.

\newpage
\begin{figure*}[htb]
\begin{center}
\hspace*{1cm}(b) ${\mathbf{q}}=\pi(1,0,0)\xi$\\[2ex]
\end{center}
\leavevmode \epsfclipon \epsfxsize=7.9truecm
\vbox{\epsfbox{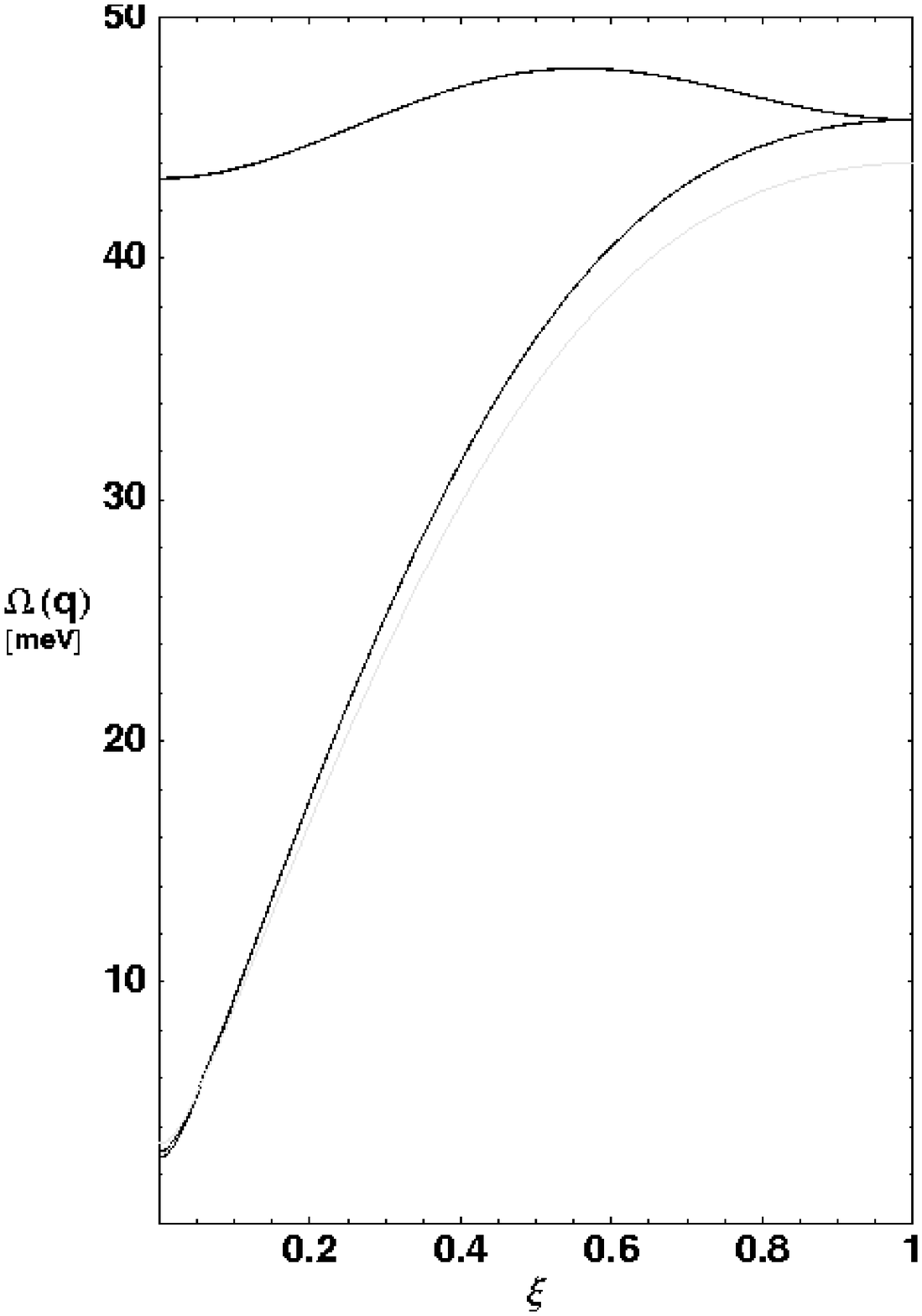}} \vspace{0.5cm}
\end{figure*}

In contrast to the absence of experimental evidence for the
optical modes in the neutron scattering experiment, Raman
spectroscopy \cite{grueninger} at low temperatures does show a
pronounced peak centered at about 40\,meV. This energy is
consistent with our  calculated optical branches
$\Omega_3({\mathbf{q}})$ and $\Omega_4({\mathbf{q}})$. In Raman spectroscopy only the zero wave vector excitation of the optical branches can be observed.
In principle, Raman spectroscopy is only sensitive to $S_z$=0 excitations
but this selection rule can be broken by the spin-orbit coupling.
 The Raman peak
disappears at the N$\acute{\mbox{e}}$el temperature, giving evidence
for a magnetic origin.  Studying spin-wave energies in Raman spectroscopy might
be subject to similar difficulties as neutron scattering when it
comes to the phonons' role.  Since the pronounced peak at about
40\,meV has a very large intensity, its explanation may
well have to include the coupling to lattice modes,  in
addition to the optical spin-wave modes.

\newpage
\begin{figure*}[htb]
\begin{center}
\hspace*{1cm}(c) ${\mathbf{q}}=\frac{\pi}{2}(0,0,1)\xi$\\[2ex]
\end{center}
\leavevmode \epsfclipon \epsfxsize=7.9truecm
\vbox{\epsfbox{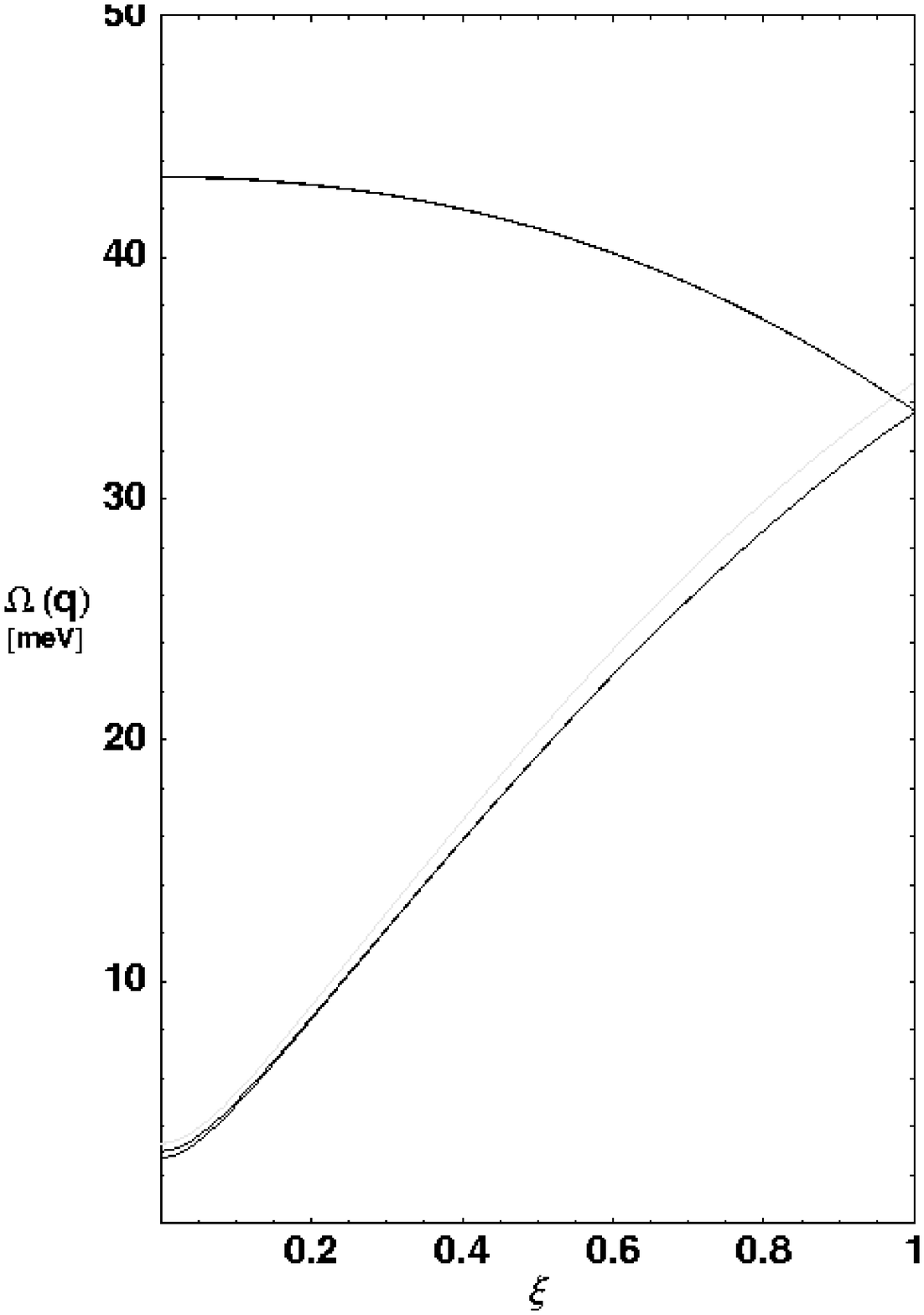}} \vspace{0.5cm} 
\end{figure*}

\section{Summary}

We have presented a detailed analysis of the spin-wave spectrum
in LaTiO$_3$. We have found that the spin-wave spectrum of this
system consists of two pairs of quasi-degenerate branches. The
modes belonging to one of the pairs have a rather small
zone-center gap, about 3\,meV, and are approximately isotropic
over the Brillouin zone. The dispersion and the gap of these two
modes are shown to reproduce the experimental data of the neutron
scattering experiment carried out on LaTiO$_3$. \cite{keimer} The
quasi-degenerate modes belonging to the second pair have a large
zone-center gap, about 43\,meV, and their dispersion shows sizeable
tetragonal anisotropy in the Brillouin zone. While not yet
detected in neutron scattering experiments, (perhaps for technical
reasons as indicated above), the zone-center gap of these modes is
consistent with Raman data. \cite{grueninger}

Our spin-wave dispersion is calculated on the basis of the
detailed low-temperature magnetic structure of LaTiO$_3$, which we
have analyzed in a previous paper. \cite{us} There, we have used
the experimentally verified orbital ordering in this system, to
develop the superexchange interaction between nearest-neighbor Ti
ions. As detailed in Ref.~\onlinecite{us}, and summarized in the
Introduction section above, the complicated magnetic structure that
we have obtained, which involves a predominant G-type
antiferromagnetic order along the $a$ axis and a canted
ferromagnetic one along the $c$ axis, agrees beautifully with all
available experimental findings. In view of the good agreement we
have found in the present study with the neutron and Raman
scattering data, it might be concluded that our analysis has
yielded a detailed understanding of the magnetism in LaTiO$_3$. In
addition, we have indicated above a rather detailed prediction
regarding the behavior of the higher-energy modes. We hope that
these will be studied experimentally, and will be compared with
our calculations.

\acknowledgements

We gratefully acknowledge discussions with M. Gr\"uninger and B.
Keimer. This work was partially supported by the German-Israeli
Foundation for Research (GIF), and by the US-Israel Binational
Science Foundation (BSF).

\end{multicols}

%\newpage
\appendix
\begin{multicols}{2}

\section{the choice of the rotation matrix} \label{explcoordrot}

As is mentioned in the text, the local coordinate system given in
Eq.~(\ref{localcoor}), in which the local $z$ axis points along
the direction of the moment in the classical ground state is
still ambiguous in that it can be rotated arbitrarily
around its $z$-axis. Here we show that our choice, Eqs.
(\ref{localcoor}) and (\ref{Ui}), leads to a considerable
simplification in the calculation of the spin-wave Hamiltonian.

Let us suppose that the local coordinate system of Eq.
(\ref{localcoor}) at each of the four lattices sites is further
rotated around its local $z$-axis by an angle $\rho_{i}$,
($i=1,2,3,4$). The rotation matrices $U_{i}$ of Eq.~(\ref{Ui}) are
then replaced by
\begin{equation}
U_{i\rho_i}=U_i \cdot \left[\begin{array}{ccc}
\,\cos \rho_i \, & -\sin \rho_i &\, \; \, 0 \, \; \, \\
\sin \rho_i \, & \;\;\,\cos \rho_i  & 0 \\
0 & 0 & 1
\end{array} \right],\label{arbr}
\end{equation}
and the corresponding superexchange matrices, Eq.~(\ref{trafo1}),
are transformed accordingly as
\begin{equation}
A'_{mn\rho_m \rho_n}=U_{m\rho_m}^t \cdot A'_{mn} \cdot U_{n\rho_n}^{\nil},
\end{equation}
where
\begin{eqnarray}
A'_{mn00}\equiv A'_{mn}
\end{eqnarray}
is the superexchange matrix of Eq.~(\ref{trafo1}).

The arbitrary rotations described above will modify the
coefficients $C_{mn}(\ell )$, Eqs.~(\ref{BosSBHam1}), appearing in
the spin-wave Hamiltonian. Denoting these modified coefficients by
$C_{mn\rho_m \rho_n}(\ell )$, such that $C_{mn00}(\ell )\equiv
C_{mn}(\ell )$, we find  the following inter-relations (using the
symmetries listed in Table \ref{macsym})
\end{multicols}
\begin{eqnarray}
C_{12\rho_1\rho_2}^{\nil}(1)&= &C_{34\rho_3\rho_4}^{\nil}(1)=
C_{16\rho_1\rho_2}^{\nil}(1)=C_{38\rho_3\rho_4}^{\nil}(1)=
C_{12}^{\nil}(1)=C_{12}^*(1), \nonumber \\
C_{13\rho_1\rho_3}^{\nil}(1)&=
&C_{24\rho_2\rho_4}^{\nil}(1)=C_{13}^{\nil}(1)
=C_{13}^{*}(1), \nonumber \\
C_{12[16]\rho_1\rho_2}^{\nil}(2)&=
&C_{12[16]}^{\nil}(2)e^{i(\rho_2-\rho_1)}, \quad
C_{34[38]\rho_3\rho_4}^{\nil}(2)
=C_{34[38]}^{\nil}(2)e^{i(\rho_4-\rho_3)},  \nonumber \\
C_{12\rho_1\rho_2}^{\nil}(2)&= &C_{16\rho_1\rho_2}^{\nil}(2)=
C_{34\rho_3\rho_4}^{\nil}(2) = C_{38\rho_3\rho_4}^{\nil}(2) \text{
if } \rho_4-\rho_3
=\rho_2-\rho_1, \nonumber \\
C_{13\rho_1\rho_3}^{\nil}(2)&=&C_{24\rho_2\rho_4}^{\nil}(2)=
C_{13}^{\nil}(2)=C_{13}^{*}(2)\nonumber \\
C_{12[16]\rho_1\rho_2}^{\nil}(3)&=&C_{12[16]}^{\nil}(3)e^{-i(\rho_1+
\rho_2)}, \quad   C_{34[38]\rho_3\rho_4}^{\nil}(3)=
C_{34[38]}^{\nil}(3)e^{-i(\rho_3+\rho_4)}\nonumber \\
C_{12\rho_1\rho_2}^{\nil}(3)&=&C_{16\rho_1\rho_2}^{*}(3)=
C_{34\rho_3\rho_4}^{\nil}(3) =C_{38\rho_3\rho_4}^{*}(3) \text{ if
} \rho_1+\rho_2=
\rho_3+\rho_4, \nonumber \\
C_{13\rho_1\rho_3}^{\nil}(3)&=&C_{13}^{\nil}(3)e^{-i(\rho_1+\rho_3)},
 \quad C_{24\rho_2\rho_4}^{\nil}(3)=C_{24}^{\nil}(3)
 e^{-i(\rho_2+\rho_4)},\quad
C_{13}^{\nil}(3)=C_{24}^{*}(3). \label{coeffsymmetries}
\end{eqnarray}
It is thus seen that with the choice employed in Eq.
(\ref{localcoor}), namely $\rho_i=0$ ($i=1,2,3,4$), the following
relations are obtained
\begin{eqnarray}
C_{12}(1)&=&C_{16}(1 )=C_{34}(1)=C_{38}(1),\ \ \
C_{13}(1)=C_{24}(1),\ \ \ 2C_{12}(1)+C_{13}(1)\equiv\frac{1}{2}
C_{1},\nonumber\\
C_{12}(2)&=&C_{16}(2)=C_{34}(2)=C_{38}(2)\equiv
\frac{1}{2}C^{\parallel}_{2},\ \ \ C_{13}(2)=C_{24}(2)\equiv
\frac{1}{2}C^{\perp}_{2}
=\frac{1}{2}C^{\perp *}_{2}\nonumber\\
C_{12}(3)&=&C_{34}(3)=C_{16}^{\ast}(3)
=C_{38}^{\ast}(3)\equiv\frac{1}{2}C^{\parallel}_{3}, \ \ \ \
C_{13}(3)=C_{24}^{\ast}(3)\equiv \frac{1}{2}C^{\perp}_{3}.
\label{gensym}
\end{eqnarray}
Here we have introduced the coefficients $C_{1}$,
$C_{2}^{\parallel,\perp}$, and $C_{3}^{\parallel,\perp}$ that are
used in our spin-wave Hamiltonian, Eq.~(\ref{start}).

As is mentioned in the text,  the Holstein-Primakoff
transformation  gives rise to terms linear in the boson operators.
The coefficients of these terms are
\begin{eqnarray}
C_{mn}(4)=\frac{1}{4}\bigl ((A'_{mn})^{xz}+i(A'_{mn})^{yz}\bigr ).
\end{eqnarray}
When summed over all single-bond contributions, these coefficients
vanish.  For example, the absolute value of the coefficient of the
boson operator $a_{\mathbf{R}}^{\dag}$ is
\begin{eqnarray}
&&2\big| C_{12}^{\nil}(4) + C_{16}^{\nil}(4)+
 C_{13}^{\nil}(4)\big|
=\mbox{$\frac{1}{2}$} \sqrt{[{A_{12}'}^{\!xz}+{A_{16}'}^{\!xz}
+{A_{13}'}^{\!xz}]^2+[{A_{12}'}^{\!yz}
+{A_{16}'}^{\!yz}+{A_{13}'}^{\!yz}]^2}=0.
\end{eqnarray}
Employing Eqs.~(\ref{trafo1}), we have written each of the terms
appearing in the square root explicitly, and verified that they
both vanish. A similar argument prevails for the other
coefficients of the linear terms.

\section{The spin-wave dispersion}
\label{diag}

In order to obtain the spin-wave dispersion resulting from the
Hamiltonian (\ref{start}), it is convenient to first introduce
a short-hand notation for this Hamiltonian. To this end we write
\begin{eqnarray}
h_{\rm SW}({\bf q})&=&\sum_{\mu\nu}\Bigl ({\cal
H}^{a}_{\mu\nu}({\bf q})\xi^{\dagger}_{\mu}({\bf
q})\xi^{\nil}_{\nu}({\bf q}) +\frac{1}{2}{\cal
H}^{b}_{\mu\nu}({\bf q})\xi^{\dagger}_{\mu}({\bf
q})\xi_{\nu}^{\dagger}(-{\bf q}) +\frac{1}{2}{\cal
H}^{b\ast}_{\mu\nu}({\bf q})\xi^{\nil}_{\mu}({\bf
q})\xi^{\nil}_{\nu}(-{\bf q})\Bigr ),\label{hamil}
\end{eqnarray}
where
\begin{eqnarray}
\xiv({\bf q})&=&\left [\begin{array}{c}a^{\nil}_{\bf q}\\
b^{\nil}_{\bf q}\\
c^{\nil}_{\bf q}
\\d^{\nil}_{\bf q}
\end{array}\right ],\ \ \xiv^{\dagger}({\bf q})=\left
[\begin{array}{cccc}a^{\dagger}_{\bf q},&b^{\dagger}_{\bf q},&
c^{\dagger}_{\bf q},& d^{\dagger}_{{\bf q}}\end{array}\right ],
\end{eqnarray}
and the Hamiltonian matrices are conveniently written in the form
\begin{eqnarray}
{\cal H}^{a}({\bf q})&=&\left [\begin{array}{cc}{\cal H}_{1}&{\cal H}_{2}\\
{\cal H}_{2}&{\cal H}_{1}\end{array}\right ],\ \ {\cal
H}_{1}=\left [\begin{array}{cc}C_{1}&C_{2}^{\parallel}(\cos
q_{x}+\cos q_{y})\\ C_{2}^{\parallel *}(\cos q_{x}+\cos
q_{y})&C_{1}\end{array}\right ],\ \ {\cal H}_{2}=C^{\perp}_{2}\cos
q_{z}\left [\begin{array}{cc}1&0\\0&1\end{array}\right ],
\label{ha}
\end{eqnarray}
and
\begin{eqnarray}
{\cal H}^{b}({\bf q})&=&\left [\begin{array}{cc}{\cal H}_{3}&{\cal
H}_{4}\\{\cal H}_{4}&{\cal H}_{3}\end{array}\right ],\ \ {\cal
H}_{3}=(C^{\parallel}_{3}\cos q_{x}+C_{3}^{\parallel *}\cos
q_{y})\left [\begin{array}{cc}0&1\\1&0\end{array}\right ],\ \
{\cal H}_{4}=\cos q_{z}\left
[\begin{array}{cc}C^{\perp}_{3}&0\\0&C_{3}^{\perp
*}\end{array}\right ].\label{hb}
\end{eqnarray}
[Note that $C_{1}$ and $C_{2}^{\perp}$ are real, see Appendix
\ref{explcoordrot}.]

Let us now denote the boson fields in which the Hamiltonian
(\ref{start}) is diagonalized by $\tau_{\ell}({\bf q}), $ $\ell
=1,2,3,4$. These fields are related to the original ones,
$\xi_{\ell} ({\bf q})$, by the general linear transformation
\begin{eqnarray}
\tau_{\ell}({\bf q})&=&\sum_{j}P^{\nil}_{\ell j}({\bf
q})\xi^{\nil}_{j}({\bf q})-\sum_{j}Q^{\nil}_{\ell j}({\bf
q})\xi_{j}^{\dagger}(-{\bf q}),\label{tau}
\end{eqnarray}
with
\begin{eqnarray}
\sum_{j}\Bigl (P^{\nil}_{\ell j}({\bf q})P^{\ast}_{nj}({\bf
q})-Q^{\nil}_{\ell j}({\bf q})Q^{\ast}_{nj}({\bf q})\Bigr
)=\delta_{n\ell},\ \ \  \sum_{j}\Bigl (-P^{\nil}_{\ell j}({\bf
q})Q^{\nil}_{nj}(-{\bf q})+Q^{\nil}_{\ell j}({\bf
q})P^{\nil}_{nj}(-{\bf q})\Bigr )=0,\label{or}
\end{eqnarray}
for the $\tau$ fields to obey the boson commutation relations.
In order that the $\tau$ fields will represent normal modes,
they have to satisfy
\begin{eqnarray}
\Bigl [\tau^{\nil}_{\ell}({\bf q}),h_{\rm SW}({\bf q})\Bigr
]=\Omega_{\ell}({\bf q})\tau^{\nil}_{\ell}({\bf q}),\label{em}
\end{eqnarray}
where $\Omega_{\ell}({\bf q}),$ $\ell =1,2,3,4$ are the
eigenfrequencies of our spin-wave Hamiltonian. Inserting Eqs.
(\ref{tau}) into Eq.~(\ref{em}), and equating the coefficients of
$\xi^{\nil}$ and $\xi^{\dagger}$ on both sides, we obtain
\begin{eqnarray}
\Omega_{\ell}({\bf q})P^{\nil}_{\ell j}({\bf q})&=&\sum_{n}\Bigl
(P_{\ell n}^{\nil}({\bf q}){\cal H}^{a}_{nj}({\bf q})+Q_{\ell
n}^{\nil}({\bf q}){\cal H}^{b\ast}_{nj}({\bf
q})\Bigr ),\nonumber\\
-\Omega_{\ell}({\bf q})Q_{\ell j}^{\nil}({\bf q})&=&\sum_{n}\Bigl
(Q_{\ell n}^{\nil}({\bf q}){\cal H}^{a\ast}_{nj}({\bf q})+P_{\ell
n}^{\nil}({\bf q}){\cal H}^{b}_{nj}({\bf q})\Bigr ).
\end{eqnarray}
Identifying $P^{\nil}_{\ell j}\equiv v^{\ell}_{j}$ as ``vector
number $\ell$ whose entries are $j$", and similarly for $Q_{\ell
j}\equiv u^{\ell}_{j} $ we arrive at the equations
\begin{eqnarray}
\Omega_{\ell}v^{\ell}&=&{\cal H}^{a\ast}v^{\ell}+{\cal
H}^{b\ast}u^{\ell},\ \ \ -\Omega_{\ell}u^{\ell}={\cal
H}^{a}u^{\ell}+{\cal H}^{b}v^{\ell},\label{uv}
\end{eqnarray}
where we have dropped the explicit ${\bf q}$ dependence
for brevity. From the first of Eqs.~(\ref{or}), we have
\begin{eqnarray}
|v^{\ell}|^{2}-|u^{\ell}|^{2}=1,\label{or1}
\end{eqnarray}
where $u^{\ell}$ and $v^{\ell}$ are 4-dimensional vectors.

We split the 4-dimensional vectors $u^{\ell}$ and
$v^{\ell}$ into two 2-dimensional vectors, $u^{\ell}=(u^{\ell}_{1},
u^{\ell}_{2})$, $v^{\ell}=(v^{\ell}_{1}, v^{\ell}_{2})$, and write
explicitly Eqs.~(\ref{uv}), using the definitions (\ref{ha}) and
(\ref{hb}). The resulting equations may be arranged in the form
\begin{eqnarray}
-\Omega (u^{\nil}_{1}-u^{\nil}_{2})&=&({\cal H}_{1}^{\nil}-{\cal
H}^{\nil}_{2})(u^{\nil}_{1}-u^{\nil}_{2}) +({\cal
H}^{\nil}_{3}-{\cal
H}^{\nil}_{4})(v^{\nil}_{1}-v^{\nil}_{2}),\nonumber\\
\Omega (v^{\nil}_{1}-v^{\nil}_{2})&=&({\cal H}_{1}^{\ast}-{\cal
H}^{\nil}_{2})(v^{\nil}_{1}-v^{\nil}_{2}) +({\cal
H}_{3}^{\ast}-{\cal
H}_{4}^{\ast})(u^{\nil}_{1}-u^{\nil}_{2}),\nonumber\\
-\Omega (u^{\nil}_{1}+u^{\nil}_{2})&=&({\cal H}^{\nil}_{1}+{\cal
H}^{\nil}_{2})(u^{\nil}_{1}+u^{\nil}_{2}) +({\cal
H}^{\nil}_{3}+{\cal
H}^{\nil}_{4})(v^{\nil}_{1}+v^{\nil}_{2}),\nonumber\\
\Omega (v^{\nil}_{1}+v^{\nil}_{2})&=&({\cal H}_{1}^{\ast}+{\cal
H}^{\nil}_{2})(v^{\nil}_{1}+v^{\nil}_{2}) +({\cal
H}_{3}^{\ast}+{\cal H}_{4}^{\ast})(u^{\nil}_{1}+u^{\nil}_{2}),
\end{eqnarray}
where we have also dropped the index $\ell$ for brevity. It is thus
seen that there are two types of solutions: Either
$u^{\nil}_{1}=u^{\nil}_{2}$ and $v^{\nil}_{1}=v^{\nil}_{2}$, in
which case the first couple of equations is trivially satisfied,
and it is needed to solve just the second pair of equations, or
{\it vice versa:} $u^{\nil}_{1}=-u^{\nil}_{2}$ and
$v^{\nil}_{1}=-v^{\nil}_{2}$ and then the first pair of equations
has to be solved. However, the only difference between the first
pair of equations and the second one are the signs appearing in
front of ${\cal H}_{2}$ and ${\cal H}_{4}$. Glancing at Eqs.
(\ref{ha}) and (\ref{hb}) reveals that these signs are determined
just by $\cos q_{z}$. Therefore, it suffices to solve one pair of
equations, and the solution of the second is obtained by simply changing
the sign of $\cos q_{z}$. Focusing on the first option, we find
that two of the eigenfrequencies are determined by
\begin{eqnarray}
{\rm det}\left [ \begin{array}{cccc}C_{1}+C_{2}^{\perp}\cos
q_{z}+\Omega &C^{\parallel}_{2}(\cos q_{x}+\cos
q_{y})&C^{\perp}_{3}\cos q_{z}&C_{3}^{\parallel}\cos
q_{x}+C^{\parallel *}_{3}\cos q_{y}\\
C^{\parallel *}_{2}(\cos q_{x}+\cos q_{y})&C_{1}+C_{2}^{\perp}\cos
q_{z}+\Omega &C_{3}^{\parallel}\cos q_{x}+C^{\parallel *}_{3}\cos
q_{y}&C^{\perp *}_{3}\cos q_{z}\\
C^{\perp *}_{3}\cos q_{z}&C_{3}^{\parallel *}\cos
q_{x}+C^{\parallel }_{3}\cos q_{y}&C_{1}+C_{2}^{\perp}\cos
q_{z}-\Omega &C^{\parallel *}_{2}(\cos q_{x}+\cos q_{y})\\
C_{3}^{\parallel *}\cos q_{x}+C^{\parallel }_{3}\cos
q_{y}&C^{\perp }_{3}\cos q_{z}&C^{\parallel}_{2}(\cos q_{x}+\cos
q_{y})&C_{1}+C^{\perp}_{2}\cos q_{z}-\Omega\end{array}\right ]=0. \label{det}
\end{eqnarray}
\begin{multicols}{2}

The modes
 $\Omega_{2}({\bf q})$ and $\Omega_{4}({\bf q})$
[see Eqs.~(\ref{omegas})] are the positive roots of the fourth-order polynomial in $\Omega$ given by Eq.~(\ref{det}). The other two eigenfrequencies are
found by changing the sign of $\cos q_{z}$.

%\newpage

%\begin{multicols}{2}

\setlength{\tabcolsep}{0.05cm}
\renewcommand{\arraystretch}{1.3}
\begin{table}
\caption{The  single-bond spin-exchange couplings (in meV). The
symmetric anisotropies are given as
${\mathbf{A}}_{mn}^{\text{d}}=(A_{mn}^{xx},A_{mn}^{yy},A_{mn}^{zz})$
and
${\mathbf{A}}_{mn}^{\text{od}}=(A_{mn}^{yz},A_{mn}^{xz},A_{mn}^{xy})$
for the diagonal and off-diagonal entries, respectively. }
\begin{tabular}{c}
Heisenberg couplings \\ \hline $J_{12}=17.094,\,J_{13}=13.484$ \\
\hline Moriya vectors \\ \hline ${\mathbf{D}}_{12}=(2.260, -0.884,
-0.893), \,{\mathbf{D}}_{13}=(-2.207, 0.377,0)$\\ \hline Symmetric
anisotropies\\ \hline ${\mathbf{A}}_{12}^{\text{d}}=(0.131,0,0),
\,{\mathbf{A}}_{13}^{\text{d}}=(-0.027,0,0)$,\\
${\mathbf{A}}_{12}^{\text{od}}=(0,-0.077,-0.061),
\,{\mathbf{A}}_{13}^{\text{od}}=(0,0,-0.052)$
\end{tabular}
\label{microscres}
\end{table}

\setlength{\tabcolsep}{0.05cm}
\renewcommand{\arraystretch}{1.4}
\begin{table}
\caption{The macroscopic couplings of the sublattice
magnetizations in terms of the microscopic single-bond spin
couplings. For instance,  $I_{12}=J_{12}$ but $I_{13}=J_{13}/2$,
because the coordination number of a Ti ion is 4 in the planes and
2 between the planes.}
%\begin{indented}
\begin{tabular}{c}
Isotropic couplings \\ \hline
$I_{12}=J_{12},\,I_{13}=\frac{1}{2}J_{13}$ \\ \hline
Dzyaloshinskii vectors \\ \hline
${\mathbf{D}}_{12}^{\text{D}}=(0,D_{12}^y,D_{12}^z),\,
{\mathbf{D}}_{13}^{\text{D}}=\frac{1}{2}{\mathbf{D}}_{13}^{\nil}$
\\ \hline Macroscopic symmetric anisotropies \\ \hline
$\Gamma_{12}^{\text{d}}={\mathbf{A}}_{12}^{\text{d}},\,
\Gamma_{12}^{\text{od}}=(A_{12}^{yz},0,0),\,\Gamma_{13}^{\nil}=\frac{1}{2}A_{13}^{\nil}$
\end{tabular}
\label{macrmicr}
\end{table}
\renewcommand{\arraystretch}{1}
\setlength{\tabcolsep}{0cm}

\setlength{\tabcolsep}{0.05cm}
\renewcommand{\arraystretch}{1.3}
\begin{table}
\caption{Symmetries of the magnetic  Hamiltonian due to the space
group. The relations for the anisotropic couplings are abbreviated
as follows: $(+,+,+)_{12}=(-,+,+)_{16}$ means that
${\mathbf{D}}_{12}^{\rm D}=(-D_{16}^x,D_{16}^y,D_{16}^z)$, etc.
Due to the glide planes, the Dzyaloshinskii vectors of the planar
bonds have vanishing $x$ components, and the respective symmetric
anisotropies have vanishing $xz$ and $xy$ entries. Because of the
mirror planes, the Dzyaloshinskii vectors of the inter-planar
bonds have vanishing $z$ components and the respective symmetric
anisotropies have vanishing $yz$ and $xz$ entries. }
\begin{tabular}{c}
Isotropic couplings \\ \hline $I_{12}=I_{34},\,I_{13}=I_{24}$ \\
\hline Dzyaloshinskii vectors \\ \hline
$(0,+,+)_{12}=(0,-,+)_{34}, \,(+,+,0)_{13}=(+,-,0)_{24}$  \\
\hline Macroscopic symmetric anisotropies \\ \hline
$(+,0,0)_{12}=(-,0,0)_{34}, \,(0,0,+)_{13}=(0,0,-)_{24}$
\end{tabular}
\label{macsym}
\end{table}
\renewcommand{\arraystretch}{1}
\setlength{\tabcolsep}{0cm}

\setlength{\tabcolsep}{0.05cm}
\renewcommand{\arraystretch}{1.3}
\begin{table}
\caption{The structure of the magnetic order, characterized by the
sublattice magnetizations ${\mathbf{M}}_i$ in the classical ground
state (normalized to $M$), in terms of the canting angles $\varphi$ and $\vartheta$.
We use orthorhombic coordinates, in which the $x,y,z$ axes are
oriented along the crystallographic $a,b,c$ directions. }
\begin{tabular}{c}
$x$ components: G-type\\ \hline $-M_1^x=M_2^x=M_3^x=-M_4^x=M\cos
\varphi \cos \vartheta $\\ \hline $y$ components: A-type\\ \hline
$-M_1^y=-M_2^y=M_3^y=M_4^y=M\sin \varphi \cos \vartheta $\\ \hline
$z$ components: ferromagnetic\\ \hline
$M_1^z=M_2^z=M_3^z=M_4^z=M\sin \vartheta $ \\ \hline Calculated
values of the canting angles \\ \hline
 $\varphi=1.42^\circ,\,\vartheta=0.80^\circ$ \end{tabular}
\label{cgs}
\end{table}
\renewcommand{\arraystretch}{1}
\setlength{\tabcolsep}{0cm}

\end{multicols}


\begin{thebibliography}{99}
\bibitem{explanation}
As compared to Ref.~\onlinecite{us}, we have slightly modified
some of the numerical values employed in the calculation of the
single-bond spin couplings: The effective Ti--O charge-tranfer
energy used here is $\Delta_{\text{eff}}=6.6$\,eV instead of the
value 5.5\,eV used in Ref.~\onlinecite{us}. (Hence the
spin-exchange couplings are multiplied by $(5.5/6.6)^2=0.694$
compared to the ones of that reference. This corresponds to a
renormalization of the units in which the ground-state energy is
found, but does not affect  the minimization of that energy, and
consequently, does not modify the ground-state magnetization
configuration.) In addition, the spin-exchange couplings are
computed taking into account only systematic contributions in
powers of the spin-orbit coupling. \cite{us}
\end{thebibliography}
\end{document}